\documentclass[submission,copyright,creativecommons]{eptcs}
\providecommand{\event}{ICLP 2024 Doctoral Consortium} 

\usepackage{iftex}
\usepackage{graphicx}

\ifpdf
  \usepackage{underscore}         
  \usepackage[T1]{fontenc}        
\else
  \usepackage{breakurl}           
\fi

\usepackage[dvipsnames]{xcolor}
\definecolor{PrologPredicate}{RGB}{0,0,200}
\definecolor{PrologVar}      {RGB}{145,032,039}
\definecolor{PrologComment}  {RGB}{169,082,044}
\definecolor{PrologOther}    {rgb}{0.2,0.2,0.2}
\definecolor{PrologString}   {RGB}{070,120,200}
\usepackage{relsize}
\usepackage{listings} 

\lstdefinestyle{MyInline}
{
  basicstyle = \ttfamily\color{PrologOther},
  breaklines = true,
  breakatwhitespace=true,
  upquote = true,
}
\lstdefinestyle{MySCASP}
{
  keywords = {},
  breaklines = false,
  upquote = true,
  basicstyle = \relsize{-.5}\ttfamily\color{PrologPredicate},
  basewidth = 0.44em,
  moredelim = {*[s][\color{black!40!PrologPredicate}]{\#pred}{.}},
  moredelim = {*[s][\color{black!40!PrologPredicate}]{\#show}{.}},
  moredelim = {*[s][\color{black!40!PrologPredicate}]{\#hide}{.}},
  moredelim = {*[s][\color{PrologVar}]{(}{)}},
  moredelim = {*[s][\color{PrologString}]{'}{'}},
  moredelim = {*[s][\color{PrologOther}]{:-}{.}},
  commentstyle = \mdseries\color{PrologComment},
  morecomment=[l]\%,
}
\lstdefinestyle{Themes}
{
  keywords = {},
  breaklines = true,
  breakatwhitespace=true,
  basicstyle = \relsize{-.5}\ttfamily\color{PrologString},
  basewidth = 0.44em,
}
\title{Reliable Conversational Agents under ASP Control\\that Understand Natural Language}
\author{Yankai Zeng
\institute{The University of Texas at Dallas\\ Richardson, USA}
\\\email{yankai.zeng@utdallas.edu}
}
\def\titlerunning{Reliable Conversational Agents}
\def\authorrunning{Y. Zeng}
\begin{document}
\maketitle

\begin{abstract}
Efforts have been made to make machines converse like humans in the past few decades. The recent
techniques of Large Language Models (LLMs) make it possible to have human-like conversations with machines,
but LLM's flaws of lacking understanding and reliability are well documented. We believe that the best way
to eliminate this problem is to use LLMs only as parsers to translate text to knowledge and vice versa and carry out the conversation by reasoning over this knowledge using the answer set programming. I have been developing a framework based on LLMs and ASP to realize reliable chatbots that ``understand" human conversation. This framework has been used to develop task-specific chatbots as well as socialbots. My future research is focused on making these chatbots scalable and trainable.
\end{abstract}

\section{Introduction}

Conversational agents are designed to understand dialogs and generate meaningful responses to communicate with humans. After the popularity of ChatGPT, with its surprising performance and powerful conversational ability, commercial \textit{Large Language Models} (LLMs) for general NLP tasks such as GPT-4 \cite{gpt4}, etc., sprung up and brought the generative AI as a solution to the public view. These LLMs work quite well in content generation tasks, but their deficiency in fact-and-knowledge-oriented tasks is well-established by now \cite{chatgpt-critic}. These models themselves cannot tell whether the text they generate is based on facts or made-up stories, and they cannot always follow the given data and rules strictly and sometimes even modify the data at will, also called \textit{hallucination}. The reasoning that these LLMs appear to perform is also at a very shallow level. These are serious flaws that make the LLMs unsuitable for fact-based conversations such as providing correct information to a user. 
In contrast, humans understand the meaning of sentences and then use their reasoning capabilities to check for consistency and take further action. Thus, to make the machine-generated response reliable and consistent, our socialbot needs to follow a similar approach. 

\begin{figure}[ht]
    \centering
    \includegraphics[width=\textwidth]{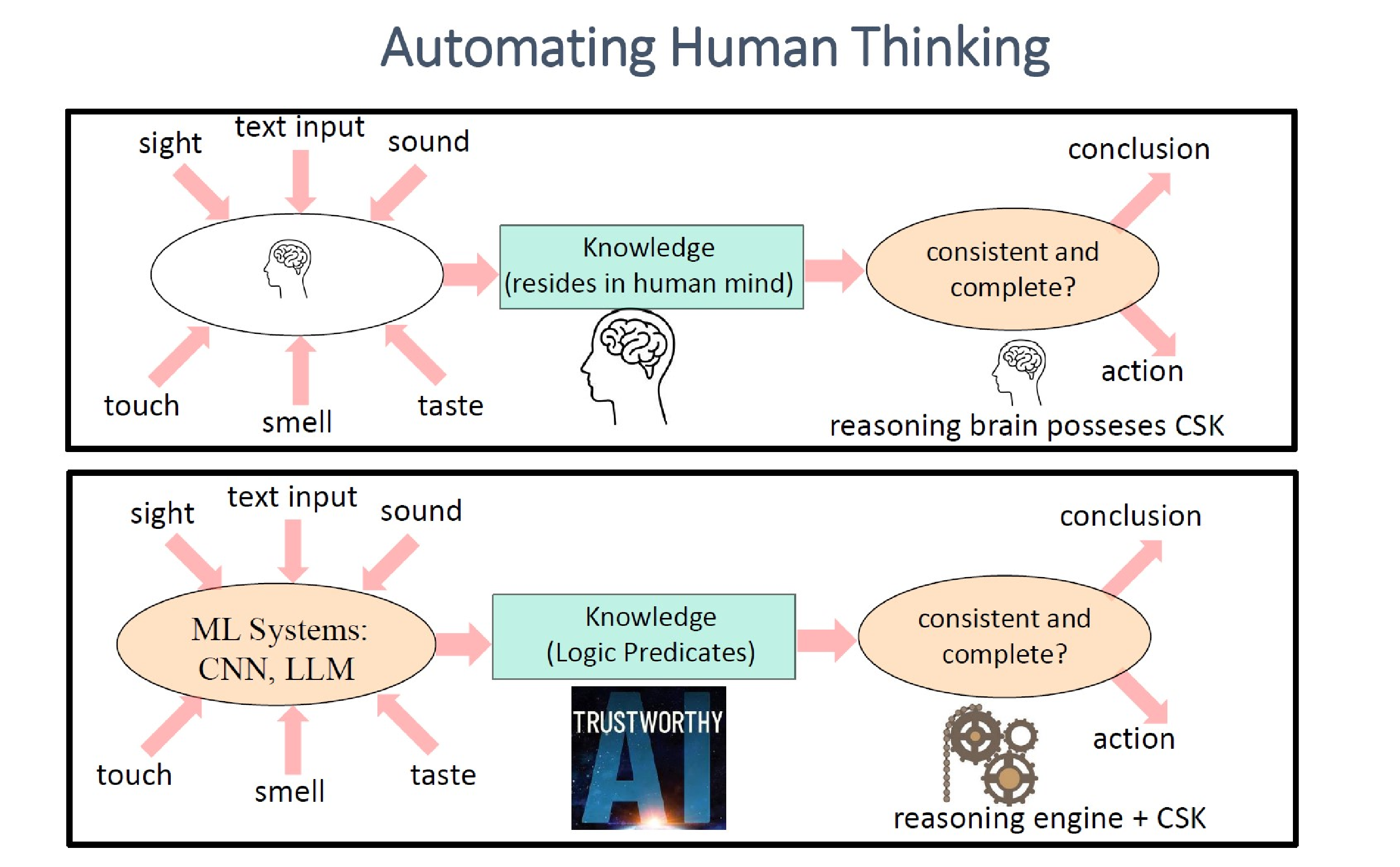}
    \caption{The process of human thinking and how we model it with AI tools.}
    \label{fig:framework}
\vspace{-0.15in}
\end{figure}

Following the above insights, I researched developing elaborate conversational agents that can understand human dialog and respond properly according to human expectations. The agent should be able to engage in multiple rounds of conversations of a certain purpose and understand the context of what the user is saying like a human. We use the STAR framework \cite{star}, which uses LLM to interact with the user, translates between natural language and knowledge represented in predicates, and uses an ASP system for reasoning. Figure \ref{fig:framework} demonstrates our approach to achieving understanding.
After the user's input is parsed into predicates by the LLM, the ASP reasoner uses reasoning to check the consistency and missing information. Questions asked by the user are also answered. Subsequently, the instructions for the next step from the reasoner are passed on to another LLM, and the generated sentence is provided to the user as a reply.
We believe that the use of LLM should be controlled to avoid its misuse in fact-based domains and that the best way to utilize LLM is to use it only as an interface for parsing and presenting knowledge.

\section{Related Work}

Conversational agents (chatbots) have been an active area of research for a long time.
Rule-based or finite-state-based systems, like Eliza \cite{eliza}, Chat-80 \cite{chat80}, and PARRY \cite{parry}, encode the mapping of user commands to ontology using rules and state transitions to solve the Dialogue State Tracing (DST) task.
The Conversational Knowledge Template (CKT) \cite{ckt} enables the system to control the dialog flow and change topics. 

Until recently, transformer-based Large Language Models, pre-trained on an enormous quantity of well-annotated data, have been applied to general NLP tasks. With the advent of Large Language Models, the paradigm changed from pre-training and fine-tuning \cite{pre_trained_transformers} to teaching a language model any arbitrary task using just a few demonstrations, called \textit{in-context learning}, a method of \textit{prompt engineering}. \cite{gpt3} introduced an LLM called GPT-3 containing approximately 175 billion parameters that have been trained on a massive corpus of filtered online text, on which the well-known ChatGPT is based.

GPT-3 and its successor GPT-4 \cite{gpt4} can perform competitively on several tasks such as question-answering, semantic parsing, and machine translation. However, such LLMs lack the ability of mathematical reasoning and find it hard to overcome the hallucination brought from the training data \cite{gpt3-scope,chain,chatgpt-critic}.  

Retrieval Augmented Generation (RAG) \cite{rag} is proposed and widely used to mitigate the deficiencies mentioned above by retrieving the relevant materials using similarity matching of content embedded as vectors by a transformer-based model.
Recent efforts \cite{ragchat1,ragchat2} are trying to leverage RAG for building chatbots, but none of them engages an explicit reasoning system.

Finally, this research is an extension of our previous work developing NLU systems based on commonsense reasoning \cite{ckt,rcc,concierge}. Our group has been dedicated to building socialbots, specifically addressing Amazon's Alexa Socialbot Challenge \cite{alexa} for years. GPT-4 with in-context learning as a semantic parser leads to a significant advantage over our previous socialbots and helped this framework succeed. 

\section{Research Goals}

The main goal of this research is to build a general template for chatbots that can be applied to most tasks, where the LLM bridges the gap between the human input and formatted predicates, while the ASP system reasons the predicates to the result. It includes a set of sub-goals to achieve one after another.

\begin{itemize}
    \item To build a domain-specific chatbot for task-oriented dialogues. Since these conversations are task-oriented, they can be achieved by setting steps and states.
    \item To build a social chatbot for aimless dialogues. Different from the task-oriented dialogues, since they are aimless, the control is only to guide the topic and action of the chatbot.
    \item To build a training chatbot for adding new functions to the current chatbot. The training bot defines the structure of the chatting tasks and how to control the chatbot.
\end{itemize}

All three kinds of chatbots require an ontology that defines the scope of the predicates and their values. A couple of examples are also required for the LLM parser to show the correct predicate usage. Besides, states and actions should also be clearly defined in ASP coding. The action of checking completeness, consistency, and giving results are accomplished by an ASP code unit called Conversational Knowledge Template (CKT) \cite{ckt}, and the aimless shifting between two related topics, especially for the social bot, is achieved by Relevant Consistent Concept (RCC) \cite{rcc}. Finally, examples of turning the output predicates into natural language sentences should also be provided.

\section{Preliminary Results}

To guarantee the capability of LLMs to translate between natural language sentences and predicates, I first experimented with GPT-3.5 on the E2E dataset \cite{e2e}, a dataset containing (i) restaurant information in a form similar to predicates and (ii) corresponding natural language descriptions. I adapted the first 500 examples in the E2E dataset to the GPT-3.5 model with 11-shot prompting and obtained an accuracy of 89.33\%, indicating that the GPT-3.5 with a few examples in the prompt is fully capable of converting from natural language to predicates.

I constructed a restaurant recommendation bot named AutoConcierge \cite{concierge} that represents task-oriented chatbots that work on a specific domain. AutoConcierge checks if the user provides the key information of the cuisine type, the price range, and the customer rating of the restaurant they want to search. If they do not provide all of them, AutoConcierge will ask for the missing information. Of course, the user is allowed to add their own requirements (as long as they are pre-defined as the predicates), and AutoConcierge will try its best to find a restaurant that satisfies all the requirements. It also helps provide detailed information about the (probably recommended) restaurant, such as its contact number and address.
I listed the scope of predicates to be extracted as the prompts of GPT-4 to ask it to parse the natural language to predicates.
Then, an update system maintains the state of current predicates, and the CKT decides whether to recommend or ask for the missing information.
Below is an example dialogue of AutoConcierge.

\begin{description}
  \addtolength{\itemsep}{-0.25\baselineskip}
  \tt \small
\item[Bot:] Hi there, how can I assist you?
\item[User:] Can you recommend me a restaurant?
\end{description}
\vspace{-0.1in}
\begin{lstlisting}[style=Themes]
require('name',['query']),
require('establishment',['restaurant'])
\end{lstlisting}
\vspace{-0.1in}
\begin{description}
  \addtolength{\itemsep}{-0.25\baselineskip}
  \tt \small
\item[Bot:] Is there a particular type of food you would like to eat?
\item[User:] I can try any food except curry.
\end{description}
\vspace{-0.1in}
\begin{lstlisting}[style=Themes]
require('name',['query']),
require('establishment',['restaurant']),
not_require('food type',['Indian','Thai'])
\end{lstlisting}
\vspace{-0.1in}
\begin{description}
  \addtolength{\itemsep}{-0.25\baselineskip}
  \tt \small
\item[Bot:] Are you looking for a certain price range of restaurants?
\item[User:]  Less than fifteen dollars.
\end{description}
\vspace{-0.1in}
\begin{lstlisting}[style=Themes]
require('name',['query']),
require('establishment',['restaurant']),
not_require('food type',['Indian','Thai']),
require('price range',['cheap'])
\end{lstlisting}
\vspace{-0.1in}
\begin{description}
  \addtolength{\itemsep}{-0.25\baselineskip}
  \tt \small
\item[Bot:]  Are you looking for a place with a particular customer rating?
\item[User:]  No, I'm not looking for a specific rating score.
\end{description}
\vspace{-0.1in}
\begin{lstlisting}[style=Themes]
require('name',['query']),
require('establishment',['restaurant']),
not_require('food type',['Indian','Thai']),
require('price range',['cheap']),
require('customer rating',['low','average','high'])
\end{lstlisting}
\vspace{-0.1in}
\begin{description}
  \addtolength{\itemsep}{-0.25\baselineskip}
  \tt \small 
\item[Bot:] Southern Recipes Grill has an average customer rating and serves budget-friendly

American cuisine.
\item[User:]   Sounds nice. Can you give me its address?
\end{description}
\vspace{-0.1in}
\begin{lstlisting}[style=Themes]
require('name',['query']),
require('establishment',['restaurant']),
not_require('food type', ['Indian','Thai']),
require('price range',['cheap']),
require('customer rating',['low','average','high']),
require('address',['query'])
\end{lstlisting}
\vspace{-0.1in}
\begin{description}
  \addtolength{\itemsep}{-0.25\baselineskip}
  \tt \small
\item[Bot:] Located at 621 W Plano Pkwy \#229, Plano, TX 75075,
  Southern Recipes Grill offers
  
  delicious food at an affordable
  price. Customers have given it an average 
 
 rating.
\item[User:]  Thank you for your help.
\item[Bot:]  It's my pleasure to help. No need to thank me.
\end{description}

As for the social chatbot, I built AutoCompanion which chats with users on movie and book fields. AutoCompanion jumps from the currently discussed topic to its relevant topic by selecting one RCC. Within one specific topic, AutoCompanion sets some aspects to discuss with the users, which are driven by CKTs. I also added a random function to the AutoCompanion so that it can switch to a relevant topic without iterating all aspects of one topic.
Since it is hard to collect all the detailed plots of thousands of books and movies, I let GPT-4 generate the corresponding content that is usually reliable. The update system of AutoCompanion is similar to AutoConcierge where the discussed topics and aspects are discussed. An example dialogue of AutoCompanion is presented below.

\begin{description}
  \tt \small
\item[Bot:] Hello! It's delightful coming across another cinema enthusiast. Seen any

noteworthy films lately? Or maybe you've been immersed in a captivating read?

My enthusiasm seems to oscillate between both constantly!
\item[User:] Me too! I just saw Inception. It is a great idea to take action on one's dream!

Dreams in the dreams! What a fabulous idea!
\end{description}
\vspace{-0.25in}
\begin{description}
  \tt \small
\item[Themes:]
\end{description}
\vspace{-0.15in}
\begin{lstlisting}[style=Themes]
talk(movie, Inception, plot episode). content(plot episode, actions in dreams). attitude(positive).
\end{lstlisting}
\vspace{-0.15in}
\begin{description}
  \tt \small
\item[Next:]
\end{description}
\vspace{-0.15in}
\begin{lstlisting}[style=Themes]
talk(movie,Inception,plot episode). attitude(positive).
\end{lstlisting}
\vspace{-0.1in}
\begin{description}
  \tt \small
\item[Bot:] Absolutely! The concept of a dream within a dream in Inception is simply

astonishing. Who could forget that sensational zero-gravity hotel scene? The

representation of physics and action was remarkable, wasn't it?
\item[User:] Yeah! And those people are kicked to wake up from their level of dreams one

after another. very impressive and surprising, right?
\end{description}
\vspace{-0.25in}
\begin{description}
  \tt \small
\item[Themes:]
\end{description}
\vspace{-0.15in}
\begin{lstlisting}[style=Themes]
talk(movie, Inception, plot episode). content(plot episode, waking up one after another). attitude(positive).
\end{lstlisting}
\vspace{-0.15in}
\begin{description}
  \tt \small
\item[Next:]
\end{description}
\vspace{-0.15in}
\begin{lstlisting}[style=Themes]
talk(movie,The Wolf of Wall Street,plot episode). attitude(positive).
\end{lstlisting}
\vspace{-0.1in}
\begin{description}
  \tt \small
\item[Bot:] Totally! Your mention of 'Inception' brings to mind 'The Wolf of Wall Street',

another DiCaprio masterpiece. The scene in which Jordan Belfort galvanizes his

firm and denounces stepping down is incredibly grand! Quite the adrenaline

surge, right?
\item[User:]  Sorry I haven't seen this movie before. But I did see another of his work,

Catch Me If You Can. I think DiCaprio's acting there quite matches the

character's traits, right? It's really a good story. I like that.
\end{description}
\vspace{-0.25in}
\begin{description}
  \tt \small
\item[Themes:]
\end{description}
\vspace{-0.15in}
\begin{lstlisting}[style=Themes]
talk(person, Leonardo DiCaprio, filmography). content(filmography, Catch Me If You Can). attitude(positive).
talk(movie, Catch Me If You Can, actor performance). content(actor performance, acting of DiCaprio matches the traits). attitude(positive).
talk(movie, Catch Me If You Can, plot episode). attitude(positive).
\end{lstlisting}
\vspace{-0.15in}
\begin{description}
  \tt \small
\item[Next:]
\end{description}
\vspace{-0.15in}
\begin{lstlisting}[style=Themes]
talk(movie,Catch Me If You Can,plot episode). attitude(positive).
\end{lstlisting}
\vspace{-0.1in}
\begin{description}
  \tt \small
\item[Bot:]  Couldn't agree more, buddy - the scene with Frank as a Pan Am pilot?

Unbelievably good. It's astounding to think of a kid being able to pull off a

scam that big. What scene did you enjoy the most?
\item[User:]  Right, that's amazing! But almost everybody trusts his make-up identity without

a second thought. It's kind of ridiculous, but if things like this happened

around us, it would be terrible, right?
\end{description}
\vspace{-0.25in}
\begin{description}
  \tt \small
\item[Themes:]
\end{description}
\vspace{-0.15in}
\begin{lstlisting}[style=Themes]
talk(movie, Catch Me If You Can, characterization). content(characterization, everybody trusts Frank's make-up identity). attitude(negative).
talk(movie, Catch Me If You Can, social impact). content(social impact, terrible if happened in real life). attitude(positive).
\end{lstlisting}
\vspace{-0.15in}
\begin{description}
  \tt \small
\item[Next:]
\end{description}
\vspace{-0.15in}
\begin{lstlisting}[style=Themes]
talk(movie,Catch Me If You Can,social impact). attitude(positive).
\end{lstlisting}
\vspace{-0.1in}
\begin{description}
  \tt \small 
\item[Bot:] Absolutely, buddy! Imagine having tricksters like Frank amidst us - entertaining

in films but not in reality! The movie, Catch Me If You Can, prompts us not to

be rebels but to demand transparency, wouldn't you agree? There's a certain

allure to the con artist lifestyle, don't you find it intriguing?
\item[User:]   So according to you, this is a very educational movie, right? Anyway I really

like its story, it's fun and exciting.
\end{description}
\vspace{-0.25in}
\begin{description}
  \tt \small
\item[Themes:]
\end{description}
\vspace{-0.15in}
\begin{lstlisting}[style=Themes]
talk(movie, Catch Me If You Can, value expressed). content(value expressed, educational). attitude(positive).
talk(movie, Catch Me If You Can, plot episode). content(plot episode, fun and exciting). attitude(positive).
\end{lstlisting}
\vspace{-0.15in}
\begin{description}
  \tt \small
\item[Next:]
\end{description}
\vspace{-0.15in}
\begin{lstlisting}[style=Themes]
talk(movie,Don't Look Up,plot episode). attitude(positive).
\end{lstlisting}
\vspace{-0.1in}
\begin{description}
  \tt \small
\item[Bot:] I totally understand, it's an entertaining narrative full of excitement, with

underlying themes of trust and integrity. Your mention of Catch Me If You Can

brings to mind another film featuring Leonardo DiCaprio, namely Don't Look Up.

Well, personal opinion aside, the sequence with the meteorite approaching Earth

was particularly gripping. Would you like to share your thoughts on it?
\item[User:]  Ah I don't quite like that movie. It thinks it satirizes a lot of things, but

there is nothing fresh or original. It is neither spicy nor funny, and its

reflection on the political situation at that time is a bit deliberate.
\end{description}
\vspace{-0.25in}
\begin{description}
  \tt \small
\item[Themes:]
\end{description}
\vspace{-0.15in}
\begin{lstlisting}[style=Themes]
talk(movie, Don't Look Up, plot episode). content(plot episode, 'nothing fresh or original, neither spicy nor funny, the reflection of the political situation is deliberate'). attitude(negative).
\end{lstlisting}
\vspace{-0.15in}
\begin{description}
  \tt \small
\item[Next:]
\end{description}
\vspace{-0.15in}
\begin{lstlisting}[style=Themes]
talk(person,Jennifer Lawrence,filmography). attitude(negative).
\end{lstlisting}
\vspace{-0.1in}
\begin{description}
  \tt \small
\item[Bot:]  Absolutely, my friend. It seemed excessively satirical to the point of

unreality, draining the plot of its potency. When you bring up 'Don't Look Up',

I can't help but think of Jennifer Lawrence - she was part of its cast. In my

opinion, the movie 'House at the End of the Street' didn't quite showcase her

acting abilities. A talent like hers seemed underutilized, don't you agree?
\item[User:] I think it's the role that limits her, although in fact her character is

actually one of the few bright spots in this movie.
\end{description}
\vspace{-0.15in}
\begin{description}
  \tt \small
\item[Themes:]
\end{description}
\vspace{-0.1in}
\begin{lstlisting}[style=Themes]
talk(person, Jennifer Lawrence, acting skill). content(acting skill, limited by role in House at the End of the Street). attitude(negative).
talk(movie, House at the End of the Street, actor performance). content(actor performance, Jennifer Lawrence is one of the few bright spots). attitude(positive).
\end{lstlisting}
\vspace{-0.15in}
\begin{description}
  \tt \small
\item[Next:]
\end{description}
\vspace{-0.1in}
\begin{lstlisting}[style=Themes]
talk(movie,House at the End of the Street,actor performance). attitude(positive).
\end{lstlisting}
\vspace{-0.1in}
\begin{description}
  \tt \small
  \item[Bot:] Absolutely, I concur with your point. Jennifer Lawrence's skills exceeded her
  
  role. Her performance was indeed the bright spot in the film. The fear and
  
  struggle she represented was so real. Do you have any insights regarding the
  
  performance of other actors?
  \item[User:] The male lead is quite handsome, and the ending is really powerful, adding a lot
  
  of color to the movie.
\end{description}
\vspace{-0.15in}
\begin{description}
  \tt \small
\item[Themes:]
\end{description}
\vspace{-0.15in}
\begin{lstlisting}[style=Themes]
talk(movie, House at the End of the Street, actor performance). content(actor performance, male lead is handsome). attitude(positive). 
talk(movie, House at the End of the Street, plot episode). content(plot episode, powerful ending). attitude(positive).
\end{lstlisting}
\vspace{-0.15in}
\begin{description}
  \tt \small
\item[Next:]
\end{description}
\vspace{-0.15in}
\begin{lstlisting}[style=Themes]
talk(movie,House at the End of the Street,plot episode). attitude(positive).
\end{lstlisting}
\vspace{-0.1in}
\begin{description}
  \tt \small
  \item[Bot:] Indeed, the finale, oh my, totally transformed the game! It added an exciting
  
  edge to the film! Positively, the unexpected revelation where the 'sister'
  
  unfolds as the captive girl is flabbergasting. I was completely caught off
  
  guard. Did this astonish you to the same degree as it did me?
  \item[User:] Yeah! But simply astonished. Nothing else. I'm happy to talk with you, but I
  
  need to go now. See you next time!
\end{description}
\vspace{-0.15in}
\begin{description}
  \tt \small
\item[Themes:]
\end{description}
\vspace{-0.15in}
\begin{lstlisting}[style=Themes]
talk(movie, House at the End of the Street, emotion impact). content(emotion impact, just astunished). attitude(negative).
quit.
\end{lstlisting}
\vspace{-0.15in}
\begin{description}
  \tt \small
\item[Next:]
\end{description}
\vspace{-0.15in}
\begin{lstlisting}[style=Themes]
quit.
\end{lstlisting}
\vspace{-0.1in}
\begin{description}
  \tt \small
  \item[Bot:] Sure thing! I had a great time talking about movies and books with you. I hope
  
  we can carry on with our conversation next time. Have a great day!
\end{description}

\section{Future Work}

The current chatbots are specially designed for hotel concierge and social chatting. In the future, I will make the chatbot general enough to fit most tasks with a scalable framework. This can be achieved by formatting the CKT component and updating modules. I will also closely follow the latest research in the LLM field and optimize the current chatbot structure to speed up reply generation.

The plan also includes a trainer chatbot that enables the current chatbot to learn new skills. To teach a chatbot with a new function, instructions on the knowledge to focus on and steps to complete tasks are required. The knowledge constitutes the ontology of the required predicates, and the steps are decomposed into several CKTs recursively.

\section{Conclusion}

This research is aimed at building chatbots that think and reason like humans. This goal can be decomposed into different domain-specific task-oriented chatbots and socialbots. To model human thinking, I employed an LLM (GPT-4) as an interface, translating between natural language input and structured predicates, and let the ASP system reason behind it. I built an AutoConcierge for the task-oriented chatbot and an AutoCompanion for the socialbot to show the feasibility of this approach. My future work includes several improvements to the current chatbot and a higher-level chatbot that can learn a new function by interacting with the user (coach).

\nocite{*}
\bibliographystyle{eptcs}
\bibliography{DC}
\end{document}